\documentclass[12pt,a4paper,english,superscriptaddress,aps,nofootinbib]{revtex4}
\usepackage[utf8]{inputenc}
\usepackage[T1]{fontenc}
\usepackage{amsmath,amssymb,graphicx}
\makeatletter
\usepackage{babel}
\usepackage[active]{srcltx}
\usepackage{graphicx,color}
\usepackage{subfigure}
\usepackage{changebar}
\usepackage{upgreek}
\usepackage{hyperref}
\hypersetup{colorlinks=true,urlcolor= blue,citecolor=blue,linkcolor= blue}
\usepackage{braket}
\usepackage{dsfont}
\usepackage{mathtools}
\usepackage{slashed}
\usepackage{empheq}
\usepackage{tikz}
\usepackage{tikz-feynman}
\usepackage{multirow}
\usetikzlibrary{decorations.pathmorphing}

\usepackage{graphicx}
\usepackage{amsfonts}
\usepackage{amsmath}
\newcommand{\CP}{\mathbb{C}\mathrm{P}}

\newcommand{\be}{\begin{equation}}
	\newcommand{\ee}{\end{equation}}
\newcommand{\bea}{\begin{eqnarray}}
	\newcommand{\eea}{\end{eqnarray}}

\begin{document}

\title{BPS lumps in the Nonminimal $\CP^1$ Maxwell-Chern-Simons Model}

\author{F. C. E. Lima}
\email{cleiton.estevao@ufabc.edu.br}
\affiliation{Centro de Matématica, Computação e Cognição (CMCC), Universidade Federal do ABC (UFABC), Av. dos Estados 5001, CEP 09210-580, Santo Andr\'{e}, S\~{a}o Paulo, Brazil.}

\author{I. B. Cunha}
\email{brito.iure@ufabc.edu.br}
\affiliation{Centro de Matématica, Computação e Cognição (CMCC), Universidade Federal do ABC (UFABC), Av. dos Estados 5001, CEP 09210-580, Santo Andr\'{e}, S\~{a}o Paulo, Brazil.}

\author{Aldo Vera}
\email{aldo.vera@umayor.cl}
\affiliation{N\'ucleo de Matem\'atica, F\'isica y Estad\'istica, Universidad Mayor, Avenida Manuel Montt 367, Santiago, Chile.}
\affiliation{Centro Multidisciplinario de F\'isica, Vicerrector\'ia de Investigaci\'on, Universidad Mayor, Camino La Pir\'amide 5750, Santiago, Chile.}

\begin{abstract}
We investigate self-dual radially symmetric configurations in the $\CP^1$ model coupled to a Maxwell and Chern-Simons (CS) gauge fields through nonminimal interactions. Starting from the nonlinear $O(3)$-sigma model, we explicitly construct its classical mapping to the $\CP^1$ formulation, highlighting the emergence of a local $U(1)$ gauge symmetry intrinsically associated with the Fubini-Study geometry of the target space. In the static regime, the combined effects of the Chern–Simons term and the Pauli-like nonminimal coupling modify the effective gauge connection, render the electric sector unavoidable, and give rise to magnetized and electrically polarized BPS lump configurations. By implementing the Bogomol’nyi procedure, we determine the self-interaction potential required for self-duality and derive the corresponding BPS equations. We show that the magnetic flux remains quantized and is completely fixed by the asymptotic behavior of the gauge field, even in the presence of the Chern–Simons and nonminimal couplings. A detailed asymptotic analysis further reveals that finite-energy solutions necessarily correspond to lump-like configurations in which the $\CP^1$ scalar field vanishes at spatial infinity. Numerical solutions of the BPS equations confirm that the resulting configurations are regular, spatially localized, and free of singularities, exhibiting confined magnetic flux together with a nontrivial localized electric field. These results show that the generalized $\CP^1$–Maxwell–CS model supports self-dual solitons whose internal structure is rigidly governed by the target-space geometry rather than by spontaneous symmetry breaking.
\end{abstract}

\maketitle

\newpage

\section{Introduction}

In 1975, Belavin and Polyakov proposed the study of the nonlinear $O(3)$-sigma model, in which they investigated metastable states in two-dimensional isotropic ferromagnets \cite{Belavin}. In this framework, the model describes a unit vector field $\mathbf{n}(x)\in S^2$, whose dynamic is purely geometrical and governed by an energy functional invariant under global $O(3)$ rotations. One of the fundamental features of the $O(3)$-sigma model is the existence of topologically nontrivial solutions, classified by the second homotopy group, $\pi_2(S^2)=\mathbb{Z}$, whose BPS-like properties\footnote{The BPS approach organizes energy to isolate positive-definite terms alongside a purely topological term proportional to a homotopy class in the configuration space. This decomposition leads to the Bogomol’nyi inequality, which establishes a lower bound on the energy within each topological sector. Configurations that saturate this bound satisfy first-order differential equations (i.e., the BPS equations), which are equivalent to the Euler–Lagrange equations. For further details, see Refs. \cite{Bogomolnyi,Prasad}.} are ubiquitous. Indeed, Ref. \cite{Belavin} paved the way for a broad range of subsequent studies, including, for instance, investigations of topological configurations in low-dimensional field theories \cite{Lima1,Lima2,Lima3,Lima4}.

Subsequently, in Ref. \cite{Stern}, Stern proposed a theory nonminimally coupled to a gauge field. Working in a three-dimensional spacetime, Stern aimed to describe point-like particles without spin degrees of freedom, yet carrying an appropriate magnetic moment. The works of Belavin-Polyakov \cite{Belavin} and Stern \cite{Stern} motivated several investigations of three-dimensional topological structures, for instance, studies of electromagnetic vortices announced in Refs. \cite{Lima5,Torres,PKGhosh}.

However, one highlights that the pioneers of the nonlinear $O(3)$-sigma model were Gell-Mann and L\'{e}vy \cite{Gellmann}, who introduced it to describe the Goldberger-Treiman relation for the decay rate of the charged pion, employing a strong interaction \cite{Schwinger} and a weak current formulated by Polkinghorne \cite{Polkinghorne}. Following the seminal work of Gell-Mann and L\'{e}vy, several studies have used the nonlinear sigma model (NLSM) in their analyses. For instance, one finds applications of the $O(3)$-sigma model in studies about the emergence of photons (see Ref. \cite{Motrunich}). Moreover, the stability of solitons and their implications were examined in Ref. \cite{Leese2}.

Beyond its relevant applications and implications, the NLSM admits a classical equivalence with the $\CP^1$ model \cite{Abraham,Amit,Banerjee1}, obtained by parametrizing the unit vector field $\mathbf{n}(x^\mu)\in S^2$ in terms of a two-component complex spinor, namely, $z=\begin{pmatrix} z_1 & z_2 \end{pmatrix}^{T}$, subject to the constraint $z^\dagger z = 1$, and to the local identification $z \to \mathrm{e}^{i\alpha(x)} z$. This reparametrization explicitly implements the geometric isomorphism $S^2 \simeq \CP^1$, under which the physical field $\mathbf{n}$ can be reconstructed via the Hopf map \cite{Nakahara,Govindarajan,Kurkccuoglu,Polyakov2,Freed,Cheng}. Consequently, the $O(3)$ global symmetry of the sigma model manifests itself as a $SU(2)$ global symmetry, while a local $U(1)$ gauge symmetry naturally emerges, concerning the phase redundancy of the field $z$.

The relevance of reformulating the NLSM in terms of the $\CP^1$ model, when viewed from the perspective of topological structures and BPS theory \cite{Bogomolnyi,Prasad}, is directly related to the fact that this formulation makes explicit the gauge structure responsible for the stability of nonperturbative solutions \cite{Alonso2}. Meanwhile, in the vector formulation the Belavin-Polyakov solutions presented in Ref. \cite{Belavin} appear as lumps described by harmonic maps $S^2 \to S^2$, the $\CP^1$ description will allow to reinterpret these structures as configurations associated with an effective magnetic flux of a $U(1)$ gauge field, whose curvature connects to the topological charge density.

In this framework, when considering a three-dimensional spacetime, the $\CP^1$ formulation is expected to provide a natural framework for the construction of BPS-like theories, in which the energy functional can be reorganized as a sum of squares supplemented by a purely topological term. The saturation of the Bogomol’nyi bound then becomes directly associated with the quantization of the magnetic flux and the topological charge, thereby ensuring the classical stability of the solutions. Furthermore, this approach allows a controlled investigation of how $\CP^1$–Maxwell fields modify the BPS configurations, leading to the emergence of magnetized non-topological lumps. This same formalism has recently been applied in the $SU(2)$-NLSM to construct magnetized baryonic layers \cite{Canfora:2023zmt} and superconducting multi-vortices \cite{Canfora:2024mkp,Vera:2024edv}, as well as in the Gross-Pitaevskii model allowing the finding of configurations with quantized fractional vorticity \cite{Canfora:2025qkl,Canfora:2025jqr} (see also Ref. \cite{Canfora:2026col} for the construction of BPS bumpy black holes). We will follow the same steps here.

We outlined this article into five sections. Initially, one studies the mapping of the NLSM to the $\CP^1$ formulation, highlighting the emergence of a target space endowed with a Fubini-Study-like metric in the second section. In the third section, by adopting a Nielsen-Olesen (NO) ansatz \cite{Nielsen}, we will investigate the emergence of BPS configurations and their properties. In the fourth section, one exposes the numerical solutions and discusses their physical implications. Finally, in the fifth section, we summarize our findings and conclusions.

\section{From the NSLM-Maxwell-CS formulation to the $\CP^1$ theory}

Nonlinear sigma models (NLSMs) and $\CP^{n}$ models play a central role in theoretical physics, as they provide an effective description of fields taking values on curved manifolds, elegantly capturing the dynamics of internal degrees of freedom and topological phenomena \cite{Schroers:1995he,Yang:1996ze,Yang:1998qca,Romao:2018egg,Bi,Polyakov,Brezin1,Friedan1,Zinn,Leese,Ward,Nguyen}. In particular, the $\CP^{n}$ model, and especially the $\CP^{1}$ case, is mathematically equivalent to the NLSM \cite{Polyakov2,Banerjee1}, a fact that is essential for the study of topological excitations such as solitons and instantons and that enables a deep understanding of ordered and disordered phases in quantum field theories and condensed matter systems \cite{Rajaraman}. From a physical perspective, these theories naturally arise in the description of quantum magnets \cite{Zhang,Senthil}, the quantum Hall effect \cite{Rajaraman2}, topological insulators \cite{Chatterjee}, and as low-energy effective theories in quantum chromodynamics \cite{Novikov}, where they serve as controllable laboratories for investigating confinement, dynamical symmetry breaking, and nonperturbative phenomena \cite{Haldane}. Consequently, the study of the NLSM and the $\CP^{1}$ model is not merely of mathematical interest. However, it provides crucial insights into how topology, symmetry, and collective interactions shape the behavior of real physical systems. In light of these applications, we now proceed to map the $O(3)$-sigma model onto a $\CP^{1}$ theory. For this purpose, let us consider the sigma field, viz., 
\begin{align}\label{Eq1}
\mathbf{n}(x^\mu)=(n^1,n^2,n^3) \qquad \mathrm{with} \qquad \mathbf{n}\cdot\mathbf{n}=1.
\end{align}

Within this framework, the Maxwell-CS-sigma action takes the form 
\begin{align}\label{Eq2}
S=\int d^3x\left[\frac{1}{2}\,(\nabla_\mu \mathbf{n})\cdot(\nabla^\mu \mathbf{n})-\frac{1}{4}F_{\mu\nu}F^{\mu\nu}+\frac{\kappa}{4}\,\epsilon^{\mu\nu\rho}A_\mu F_{\nu\rho}-\widetilde V(\mathbf{n})
\right],
\end{align}
where $\tilde{V}(\mathbf{n})$ is the interaction term from the O(3)-theory and $F_{\mu\nu}$ is the electromagnetic tensor, i.e., $F_{\mu\nu}=\partial_\mu A_\nu-\partial_\nu A_\mu$ and the generalized covariant derivative  is 
\begin{align}\label{Eq3}
\nabla_\mu \mathbf{n}=\partial_\mu \mathbf{n}+(eA_\mu+gG_\mu)\,\hat{\mathbf n}_3\times \mathbf{n} \qquad \mathrm{with} \qquad G_\mu=\frac{1}{2}\epsilon_{\mu\nu\rho}F^{\nu\rho}.
\end{align}
Here, $\kappa$ is the Chern--Simons coupling, whereas $g$ controls the nonminimal interaction and plays the role of an anomalous magnetic moment parameter ($G_\mu$).

To map the NLSM onto a $\CP^1$ theory, one adopts a parametrization in terms of the internal space, namely the complex matrices $\sigma^i$, such that $n^a=z^\dagger\sigma^az$, where $z$ is the $\CP^1$ field defined as 
\begin{align}\label{Eq4}
z=\begin{pmatrix} z_1\\ z_2 \end{pmatrix} \qquad \mathrm{with} \qquad z^\dagger z=1.
\end{align}

Within this construction, the matrices $\sigma^a$ are
\begin{align}
    \label{Eq5}
    \sigma^1=\begin{pmatrix} 0 & 1 \\ 1 & 0 \end{pmatrix}, \qquad \sigma^2=\begin{pmatrix} 0 & -i \\ i & 0 \end{pmatrix}, \qquad 
    \sigma^3=\begin{pmatrix} 1 & 0 \\ 0 & -1 \end{pmatrix}.
\end{align}
Note that the representation is invariant under the local transformation $z(x^\mu)\to \mathrm{e}^{i\alpha(x^\mu)}z(x^\mu)$, which anticipates the emergence of a $U(1)$ gauge symmetry.

By taking into account the identity $(\sigma^a)_{ij}(\sigma^a)_{kl}=2\delta_{il}\delta_{jk}-\delta_{ij}\delta_{kl}$, one finds that
\begin{align}\label{Eq6}
(\nabla_\mu \mathbf n)^2=4\left[(\mathcal D_\mu z)^\dagger (\mathcal D^\mu z)-(z^\dagger \mathcal D_\mu z)\,(\mathcal D^\mu z)^\dagger z\right],
\end{align}
where the NLSM constraint leads  to the relation $\partial_\mu(z^\dagger)z=-z^\dagger\partial_\mu z$.

Accordingly, one can write  the covariant derivative  as $\mathcal{D}_\mu z=(\partial_\mu -iA_\mu -igG_\mu)z$, yielding 
\begin{align}\label{Eq7}
    \frac{1}{2}(\nabla_\mu \mathbf n)^2=2\,(\mathcal D_\mu z)^\dagger(1-zz^\dagger)(\mathcal D^\mu z).
\end{align}

Meanwhile, the term  arising  from the Lagrange multiplier from the NLSM due to the  constraint stated in Eq \eqref{Eq1}, takes the form $\tilde{V}(\mathbf{n})\sim\tilde{V}(\vert z\vert)$. Thus, one arrives at the nonminimal $\CP^1$-Maxwell-CS model, viz., 
\begin{align}\label{Eq8}
S^{\CP^1}=\int d^3x\,\left[2\,(\mathcal D_\mu z)^\dagger
\bigl(1-zz^\dagger\bigr)(\mathcal D^\mu z)-\frac{1}{4}F_{\mu\nu}F^{\mu\nu}+\frac{\kappa}{4}\,\epsilon^{\mu\nu\rho}A_\mu F_{\nu\rho}-\widetilde V(|z|)\right].
\end{align}

Let us now define the $\CP^1$ field as
\begin{align}\label{Eq9}
z=\frac{1}{\sqrt{1+|u|^2}}\begin{pmatrix} 1 \\ u\end{pmatrix},
\end{align}
which leads to the action on the $\CP^1$ target space of the Fubini-Study theory \cite{Manton2}, namely,
\begin{align}\label{Eq10}
S^{\CP^1}=\int d^3x\left[\frac{2\,|\mathcal D_\mu u|^2}{(1+|u|^2)^2}
-\frac{1}{4}F_{\mu\nu}F^{\mu\nu}+\frac{\kappa}{4}\,\epsilon^{\mu\nu\rho}A_\mu F_{\nu\rho}-\tilde{V}(\vert u\vert)\right].
\end{align}
The kinetic sector is thus naturally endowed with the Fubini--Study metric, which emerge from the projection onto the physical tangent space orthogonal to the local $U(1)$ fiber, yielding the curved target-space geometry characteristic of $\CP^1$ \cite{Manton2}. Furthermore, the nonminimal interaction modifies the effective gauge connection felt by the $CP^1$ field through the dual vector $G_\mu$.

Naturally, the corresponding Euler--Lagrange equation for the matter field is
\begin{align}\label{Eq11}
\mathcal D_\mu\left[\frac{2\,\mathcal D^\mu u}{(1+|u|^2)^2}
\right]-\frac{4u^\dagger}{(1+|u|^2)^3}|\mathcal D_\mu u|^2+\frac{\partial \widetilde V}{\partial u^\dagger}=0.
\end{align}
Meanwhile, variation with respect to $A_\mu$ yields the generalized gauge-field equation, viz.,
\begin{align}\label{Eq12}
\partial_\nu F^{\nu\mu}+\frac{\kappa}{2}\,\epsilon^{\mu\nu\rho}F_{\nu\rho}=J^\mu+\frac{g}{e}\,\epsilon^{\mu\nu\rho}\partial_\nu J_\rho,
\end{align}
where the matter current is 
\begin{align}\label{Eq13}
J^\mu=\frac{2ie}{(1+|u|^2)^2}\left(u^\dagger \mathcal D^\mu u-u (\mathcal D^\mu u)^\dagger\right).
\end{align}
One highlights that the presence of the Chern--Simons term makes the gauge sector intrinsically parity-odd coupling to the electric and magnetic components already at the level of the equations of motion.
Simultaneously, the nonminimal Pauli-like coupling introduces higher-derivative gauge contributions through $G_\mu$, thereby modifying the structure of the static sector.

\section{On the BPS configurations}

\subsection{Physical characterization of the BPS configurations}

Henceforth, we are interested in static radially symmetric field configurations in the generalized $\CP^1$-Maxwell-CS theory given in Eq. \eqref{Eq13}. In pursuit of this purpose, one adopts the \textit{ans\"atze}
\begin{align}\label{Eq14}
u=f(r)e^{in\theta}, \qquad A_\theta=\frac{1}{e}[\,n-a(r)\,], \qquad A_r=0, \qquad \mathrm{and} \qquad A_0=A_0(r),
\end{align}
where $f(r)$, $a(r)$, and $A_0(r)$ are the radial field variables of the matter and gauge sectors, respectively. Moreover, $n$ denotes the winding number, $\theta$ is the angular coordinate, and $r$ is the radial coordinate.\footnote{One highlights that the winding number is a positive-integer, i.e., $n\in\mathbb{Z}_+$.}

Furthermore, to ensure regularity of the fields and finiteness of the total energy, the field variables must satisfy appropriate boundary conditions, i.e., near the origin, regularity requires
\begin{align}\label{Eq15}
f(0)=0, \qquad a(0)=n, \qquad \mathrm{and} \qquad A_0'(0)=0.
\end{align}
The condition $f(0)=0$ guarantees the regularity of the scalar field at the core, while $a(0)=n$ ensures that the angular gauge field remains nonsingular at $r=0$. Similarly, $A_0'(0)=0$ prevents a divergent electric field at the origin.

Meanwhile, at spatial infinity, one imposes
\begin{align}\label{Eq16}
f(\infty)=\beta_\infty, \qquad a(\infty)=n_\infty, \qquad \mathrm{and} \qquad A_0(\infty)=A_\infty,
\end{align}
where $n_\infty$ is an integer and $\beta_\infty$, $A_\infty$ are real constants. One highlights that these asymptotic values must be compatible with the vacuum structure.

Within this framework, it is convenient to define the electric and magnetic fields. In this case, the electric and magnetic field are $\mathrm{E}_i=F_{0i}$ and $B=F_{12}$ \cite{Jackiw}. Therefore, for the static radial ansatz, one finds that the electric and magnetic field are $\mathrm{E}_r=-A'_0(r)$ and $B=-a'(r)/er$. Thus, we conclude that the electric sector is purely radial, whereas the magnetic field remains orthogonal to the plane.

Since the nonminimal interaction is mediated by the dual field $G_\mu$, we implement the effective combinations, viz.,
\begin{align}\label{Eq17}
\overline a(r)\equiv a(r)-g\,rA_0'(r) \qquad \mathrm{and} \qquad
\overline A_0(r)\equiv eA_0(r)+gB(r).
\end{align}
In terms of these quantities, one notes that the generalized covariant derivatives acting on the $\CP^1$ field, which incorporate the nonminimal Pauli-like interaction, modify both the angular and temporal gauge couplings through the effective functions $\overline a(r)$ and $\overline{A}_0(r)$.

Once the magnetic field is $B=-a'(r)/er$, one concludes that the magnetic flux emitted is
\begin{align}\label{Eq18}
\Phi_{\mathrm{flux}}=-\frac{2\pi}{e}\int_0^\infty dr\,a'(r)=\frac{2\pi M}{e},
\end{align}
with $M=n-n_\infty\in \mathbb{Z}$. Hence, the magnetic flux is quantized even in the presence of the Chern--Simons term and the nonminimal coupling.

By considering the generalized gauge-field equation in the static regime, one obtains the radial form of Gauss's law, viz., 
\begin{align}\label{Eq19}
    \frac{1}{r}\bigl(rA_0'(r)\bigr)'+\kappa B=\frac{4ef^2}{(1+f^2)^2}\,\overline A_0-\frac{4g}{r}\left[\frac{\overline a\,f^2}{(1+f^2)^2}\right]'.
\end{align}
This equation shows that one cannot consistently impose $A_0=0$ in general. Indeed, the Chern--Simons term couples the electric and magnetic sectors, whereas the nonminimal interaction contributes an additional derivative term. 

The topological configurations in the $\CP^1$-Maxwell model are completely characterized by the adopted radial ansatz in Eq. \eqref{Eq14} and by the topological boundary conditions in Eqs. \eqref{Eq15} and \eqref{Eq16} imposed on the field variables $f(r)$ and $a(r)$. The analysis of Gauss’s law in the static regime ensures the electromagnetic nature of the topological structures. In this context, the magnetic flux emerges as a quantized topological quantity, determined exclusively by the asymptotic values of the gauge field. 

Note that while for the pure Maxwell case the $n=1$ solution is prohibited by Yang's theorem \cite{Yang:1996ze,Yang:1998qca}, the inclusion of the Chern-Simons term and the nonminimal coupling allows for regular solutions in this sector.

Consequently, the topological structure is rigidly controlled  by the asymptotic values of the gauge field, while the electric sector becomes nontrivial and must be treated on the same footing as the magnetic one. 

\subsection{The BPS approach}

Once the preliminary definitions from the $\CP^1$-Maxwell-CS model are explained, we can proceed to the study the BPS properties of the field configurations in a Fubini-Study-like target space which emerge from the $\CP^1$-Maxwell-CS theory. Toward this end, let us begin by considering the energy carried by the $\CP^1$-Maxwell-CS fields. Within this framework, the energy in the static case is\footnote{For simplicity, we adopt natural units, i.e., $\hbar=e=1$.}
\begin{align}\label{Eq20}
    E=\int\, d^2x\,\left[\frac{2|D_i u|^2}{(1+|u|^2)^2}+\frac{2|D_0 u|^2}{(1+|u|^2)^2}+\frac{1}{2}\mathrm{E}_i^2+\frac{1}{2}B^2+\widetilde V(\vert u\vert)\right],
\end{align}
with $i=1,2$. By employing the radial ansatz introduced in
Eq. \eqref{Eq14}, one notes that electrical and magnetic fields are, respectively, 
\begin{align}
    \mathrm{E}_r=-A_0'(r) \qquad \mathrm{and} \qquad B=-\frac{a'(r)}{r}.
\end{align}

Meanwhile, by reformulating the energy from the $\CP^1$-Maxwell-CS model takes the form
\begin{align}\nonumber 
    E=&\int\, d^2x\,\Bigg[\frac{2}{(1+f^2)^2}\left(f'^2+\frac{\bigl[a(r)-g\,rA_0'(r)\bigr]^2f^2}{r^2}+\bigl[A_0(r)+gB(r)\bigr]^2f^2\right)+\frac{1}{2}\left(\frac{a'(r)}{r}\right)^2+\\
    +&\frac{1}{2}A_0'(r)^2+\tilde{V}(f)\Bigg],
    \label{Eq21}
\end{align}
where $A_0(r)+gB(r)=0$.

Furthermore, let us introduce an auxiliary function $\mathcal{W}(f)$ (or superpotential), viz., 
\begin{align}
    \frac{d\mathcal{W}}{df}=2\,\Omega(f)\,\Delta(f)\,f=\frac{4f}{(1+f^2)\sqrt{(1+f^2)^2+4g^2f^2}},
\label{Eq22}
\end{align}
where
\begin{align}
    \Delta(f)\equiv\sqrt{1+\frac{4g^2f^2}{(1+f^2)^2}} \qquad \mathrm{and} \qquad \Omega(f)\equiv\frac{2}{(1+f^2)^2+4g^2f^2}.
\label{Eq23}
\end{align}
By adopting the normalization $\mathcal{W}(0)=0$, one may write the superpotential as
\begin{align}\label{Eq25}
    \mathcal{W}(f)=\frac{1}{g}\left[\arctan(g)-\arctan\left(\frac{g(1-f^2)}{(1+f^2)\sqrt{1+g^2}}\right)\right] \qquad \mathrm{with} \qquad g\neq 0.
\end{align}
By taking into account Eqs. \eqref{Eq21} and \eqref{Eq25}, we rewrite the energy as
\begin{align}\nonumber
    E&=\int\,d^2x\,\Bigg[\Omega(f)\left(f'\pm\frac{f}{r}\bigl[a(r)-g\,rA_0'(r)\bigr]\Delta(f)\right)^2+\frac{1}{2}\left(A_0'(r)\pm g\,\frac{d\mathcal{W}}{df}\,f'\right)^2\\
    &+\frac{1}{2}\left(\frac{a'(r)}{r}\pm \mathcal{W}(f)\right)^2+\frac{2f^2}{(1+f^2)^2}\bigl[A_0(r)+gB(r)\bigr]^2+\tilde{V}(f)-\frac{\mathcal{W}(f)^2}{2}\Bigg]+E_{\mathrm{BPS}},
\label{Eq28}
\end{align}
where the BPS energy is
\begin{align}
    E_{\mathrm{BPS}}=\mp\int\,d^2x\,\frac{1}{r}\frac{d}{dr}\Bigl[a(r)\,\mathcal{W}(f(r))\Bigr]=\mp 2\pi\,[\,a(r)\mathcal{W}(f(r))\,]_{r=0}^{r=\infty}.
\label{Eq29}
\end{align}

Therefore, for the model to support a BPS property, one must require
\begin{align}
\widetilde V(f)=\frac{\mathcal{W}(f)^2}{2}.
\label{Eq130}
\end{align}
Therefore, the energy announced in Eq. \eqref{Eq28} is bounded from below, i.e., $E\ge E_{\mathrm{BPS}}$. In the energy-saturation limit, namely, $E=E_{\mathrm{BPS}}$, one obtains the self-dual equations, viz.,
\begin{align}
    f'=\mp\frac{f}{r}\bigl[a(r)-g\,rA_0'(r)\bigr]\Delta(f) \qquad \mathrm{and} \qquad \frac{a'(r)}{r}=\mp \mathcal{W}(f),
\label{Eq31}
\end{align}
with $A_0'(r)=\mp\,g\,\frac{d\mathcal{W}}{df}\,f'$.

The BPS formulation in the $\CP^1$-Maxwell-CS model reveals that saturation of the lower energy bound follows from an appropriate choice of the self-interaction potential in Eq. \eqref{Eq130} (and Eq. \eqref{Eq25}) in terms of an auxiliary function $\mathcal{W}(f)$ compatible with the geometry of the Fubini-Study target space. In this framework, the BPS bound enables us to clearly identify the connections between the kinetic sector, the electrical and magnetic fields, and the effective potential, resulting in a set of first-order self-dual equations. These BPS equations not only ensure the minimization of the total energy but also guarantee the classical stability of the configurations, since the energy depends exclusively on the boundary conditions, as can be seen from Eqs. \eqref{Eq15} and \eqref{Eq16}.

\subsection{On the asymptotic behavior}\label{SecIII}

Since the boundary conditions stated in Eq. \eqref{Eq12} must be satisfied, near the core, one must have $f(0)=0$ and $a(0)=n$. Therefore, the first BPS equation leads us to 
\begin{align}\label{AB1}
    f'\simeq\mp\frac{n}{r}f,
\end{align}
whose regular solution is 
\begin{align}\label{AB2}
    f(r)\simeq\mathcal{C}_0\, r^{\vert n\vert} \qquad \mathrm{with} \qquad r\to 0 \ ,
\end{align}
where $\mathcal{C}_0$ is a real constant. One highlights that any other behavior (for instance, $f \sim r^{-\vert n\vert}$) is divergent and therefore ruled out by regularity.

Since, as $r\to 0$, one has $f\sim r^{\vert n\vert}$, by analyzing the second BPS equation in Eq. \eqref{Eq21}, we find that
\begin{align}\label{AB3}
    \frac{a'}{r}\sim \mp 2\mathcal{C}_0^2r^{2n}
\end{align}
which yields
\begin{align}\label{AB4}
    a(r)\simeq  n-\frac{\mathcal{C}_0^2}{\vert n\vert+1}r^{2\vert n\vert +2} \qquad \mathrm{with} \qquad r\to 0.
\end{align}
Similarly, by adopting the asymptotic behavior \eqref{AB2} and \eqref{AB4}, one concludes that near the origin (i.e., $r\to 0$), one obtains
\begin{align}
    A_0(r)\simeq -2g\,\mathcal{C}_0^2r^{2\vert n\vert},
\end{align}

Therefore, $f(r)$, $a(r)$ and $A_0(r)$ is regular and nearly constant near $r=0$. Naturally, the regularity of the solutions at the origin, together with the boundary conditions in Eq. \eqref{Eq12}, enforces a power-law behavior for the scalar field, ensuring finiteness of the energy and the absence of physical singularities. Consequently, the gauge field remains approximately constant near the origin, admitting a regular expansion. This behavior guarantees that the magnetic field is finite at $r=0$ and that the magnetic flux is quantized by the topological integer, i.e., the winding number.

Meanwhile, the boundary conditions in Eq. \eqref{Eq12} require that far from the vortex $f(\infty)=\beta_\infty$ and $a(\infty)= n_\infty$. Therefore, at spatial infinity one has $f(r)=\beta_{\infty}-\delta f(r)$, with $\delta f \ll 1$, which allows us to write
\begin{align}
    f(r)\simeq \frac{\mathcal{C}_\infty}{r^{\vert n_\infty\vert}} \qquad \mathrm{with} \qquad r\to \infty.
\end{align}
Simultaneously,
\begin{align}
    \frac{a'}{r}\simeq 2f^2\equiv 2\beta_\infty^2.
\end{align}
where $\beta_\infty$ is a constant. Therefore, for $a(r)$ to approach a finite value, i.e., $n_\infty$, one must necessarily have $\beta_\infty = 0$ or $\beta_\infty = \infty$. Since $\beta_\infty$ leads to divergent solutions, one concludes that $\beta_\infty=0$.\footnote{Although the BPS energy \eqref{Eq29} reduces to a boundary term, it vanishes identically due to the lump boundary conditions.} Consequently, only lump-like solutions can arise in this theory.

By assuming the asymptotic behavior $\beta_\infty = 0$, one concludes that
\begin{align}
    a(r)\simeq n_\infty+\mathcal{O}(r^{2-2\vert n_\infty\vert}) \qquad r\to \infty.
\end{align}
Therefore, in the asymptotic limit, the consistency of the BPS equations requires that the scalar field approaches an admissible vacuum value, $f(r) \to \beta_\infty$, with finite-energy solutions demanding $\beta_\infty = 0$. In this regime, the field decays resemble a power law, reflecting the massless nature of the excitations in the $\CP^1$-Maxwell-CS model. Consequently, the gauge field approaches its asymptotic value as $a(r) \simeq n_\infty - 2/r^2 + \dots$. This behavior ensures the convergence of the magnetic flux and an unambiguous definition of the topological charge of the self-dual solutions.

\section{The numerical solutions}

\subsection{On the numerical approach}\label{IVA}

To achieve our goal of studying magnetized lump solutions, it becomes necessary to examine the numerical solutions of Eq. \eqref{Eq21}, subject simultaneously to the boundary conditions explicitly stated in Eq. \eqref{Eq12}. For this purpose, we may employ the Runge-Kutta numerical method, which is efficient to examine equations of the type
\begin{align}\label{Eq30}
    \frac{df_1}{d\varsigma}=h_1(f_1,f_2;\varsigma) \hspace{1cm} \text{and} \hspace{1cm} \frac{df_2}{d\varsigma}=h_2(f_1,f_2;\varsigma) \ ,
\end{align}
with the boundary conditions displayed in Eq. \eqref{Eq12}. To solve our problem numerically, it is necessary to adopt a finite range, e.g., $\varsigma \in [\varsigma_1, \varsigma_2]$, discretized into steps of size $h$.

To obtain the numerical solutions of Eq. \eqref{Eq21}, we employed the fourth-order Runge-Kutta method, applied simultaneously to all dynamical variables. In this framework, the independent variable is $\varsigma$, while the dependent variables are $f_1$ and $f_2$ ($f_1=f$ and $f_2=a$). Accordingly, at each integration step $\varsigma \to \varsigma_{m+1} = \varsigma_m + h$, the method allows us to compute the corresponding numerical coefficients, viz.:
\begin{align}\nonumber
    & k_{1,1}=h_1(\varsigma_n, f_{1,n}, f_{2,n}) \ , \\ \nonumber
    & k_{1,2}=h_2(\varsigma_n, f_{1,n}, f_{2,n})\ , \\ \nonumber
    & k_{2,1}=h_1\left(\varsigma_n+\frac{h}{2}, f_{1,n}+\frac{h}{2}k_{1,1}, f_{2,n}+\frac{h}{2}k_{1,2}\right) \ , \\ \label{Eq31}
    & k_{2,2}=h_2\left(\varsigma_n+\frac{h}{2}, f_{1,n}+\frac{h}{2}k_{1,1}, f_{2,n}+\frac{h}{2}k_{1,2}\right) \ ,\\ \nonumber
    & k_{3,1}=h_1\left(\varsigma_n+\frac{h}{2}, f_{1,n}+\frac{h}{2}k_{2,1}, f_{2,n}+\frac{h}{2}k_{2,2}\right) \ , \\ \nonumber
    & k_{3,2}=h_2\left(\varsigma_n+\frac{h}{2}, f_{1,n}+\frac{h}{2}k_{2,1}, f_{2,n}+\frac{h}{2}k_{2,2}\right) \ , \\ \nonumber
    & k_{4,1}=h_1(\varsigma_n+h\, f_{1,n}+h\, k_{3,1}, f_{2,n}+h\, k_{3,2}) \ , \\ \nonumber
    & k_{4,2}=h_1(\varsigma_n+h\, f_{1,n}+h\, k_{3,1}, f_{2,n}+h\, k_{3,2}) \ .
\end{align}

By adopting the coefficients of the Eq. \eqref{Eq31}, we obtain the $m$-th solution, i.e.,
\begin{gather}
    \label{Eq32}
    f_{1,n+1}=f_{1,n}+\frac{h}{6}(k_{1,1}+2k_{2,1}+2k_{3,1}+k_{4,1})  \ , \\ 
    \label{Eq33}
    f_{2,n+1}= f_{2,n}+\frac{h}{6}(k_{1,2}+2k_{2,2}+2k_{3,2}+k_{4,2}) \ .
\end{gather}
To achieve our goal, we discretize the radial variable $r$ in Eq. \eqref{Eq21} as $r_1, r_2, \dots, r_{10000}$, where $r_1 = 0$ and $r_{10000} = 10$. That allows us to assign values to the field variables, namely $(r_n, f_n)$ and $(r_n, a_n)$, over the discrete interval $[0,\,10]$ with step size $h = 10^{-3}$. We then apply numerical interpolation to construct continuous numerical solutions for the field variables $f(r)$ and $a(r)$. Generally speaking, we employed low-degree polynomial functions to ensure smooth transitions between discrete points and the existence of derivatives. For more details about this approach, see Refs. \cite{Burden,Butcher,Atkinson}.

\subsection{The numerical solution}

\begin{figure}[!ht]
    \centering
    \subfigure[Numerical solution of the field variable $f(r)$.]{\includegraphics[width=7cm,height=5cm]{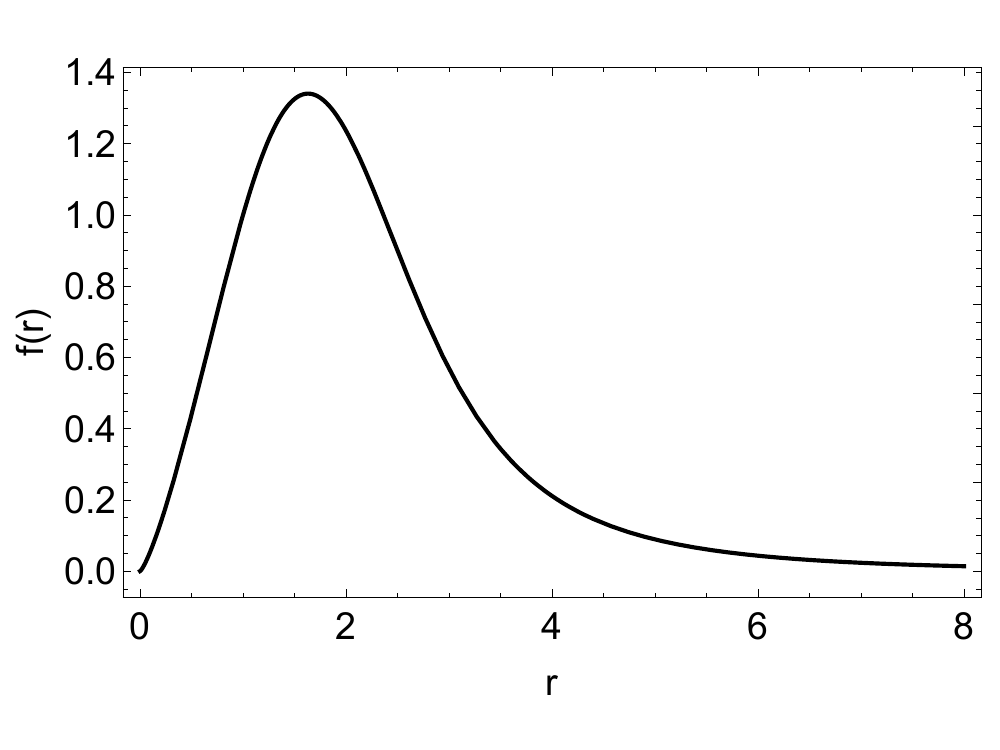}}\hfill
    \subfigure[Numerical solution of the field variable $a(r)$.]{\includegraphics[width=7cm,height=5cm]{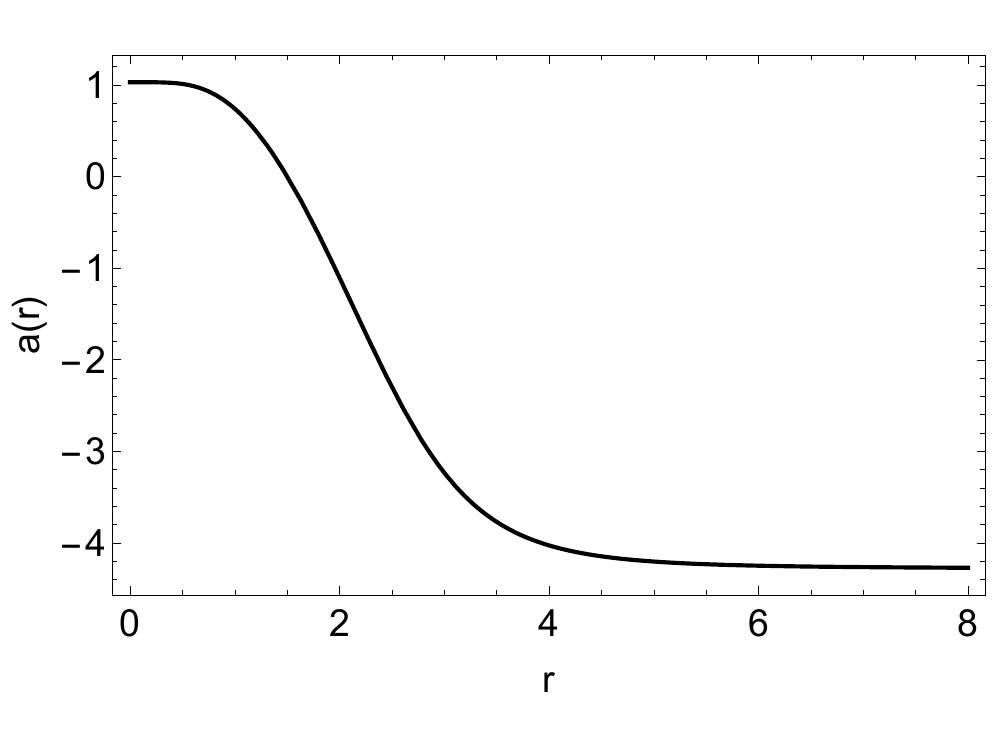}}
    \caption{Numerical solutions with $g=0.025$, $e=1$ and unit winding number.}
    \label{Fig1}
\end{figure}

By establishing the self-dual equations in Eq. \eqref{Eq21} and their asymptotic behavior (see Section \ref{SecIII}), and taking into account the numerical framework described in Section \ref{IVA} with the boundary conditions in Eq. \eqref{Eq11}, one obtains the numerical solutions. One displays the corresponding numerical results in Figs. \ref{Fig1}(a) and \ref{Fig1}(b).

Note that the numerical solutions are in agreement with the topological boundary conditions stated in Eq. \eqref{Eq11}, as well as with the asymptotic behavior detailed in Section \ref{SecIII}. That allows us to characterize the solutions as magnetized BPS lump configurations, since the $\CP^1$ field sector describes configurations satisfying $f_0\to 0$ and $f_\infty\to 0$.

In Fig. \ref{Fig1}(a), the scalar field $f(r)$ vanishes at the origin, in agreement with regularity at the structure core and with the analytical behavior $f(r)\sim r^{\vert n\vert}$ in the limit $r\to 0$. It then increases smoothly, reaching a maximum at an intermediate radial region, and subsequently decays back to zero at infinity. This profile confirms that the 
$\CP^1$ field does not interpolate between distinct vacua, returning to the same asymptotic value $f(\infty)$, thereby characterizing a lump-like solution in the $\CP^1$ target space, in contrast to the conventional NO-like vortices. Figure \ref{Fig1}(b) shows the behavior of the gauge function $a(r)$, which starts at $a(0)=n$, reflecting the unit winding number, and decreases monotonically toward a constant asymptotic value $a(\infty)=n_\infty$, ensuring magnetic flux quantization. The smoothness of $a(r)$ over the entire radial domain indicates the absence of singularities and guarantees that the associated magnetic field is finite and spatially localized. Taken together, these figures corroborate the consistency of the numerical solutions of the BPS equations with the imposed topological boundary conditions, demonstrating that the system’s energy saturates the BPS bound and is determined exclusively by the asymptotic values of the fields.

\begin{figure}[!ht]
    \centering
    \subfigure[Numerical solutions: $B(r)$ vs. $r$.]{\includegraphics[width=7cm,height=5cm]{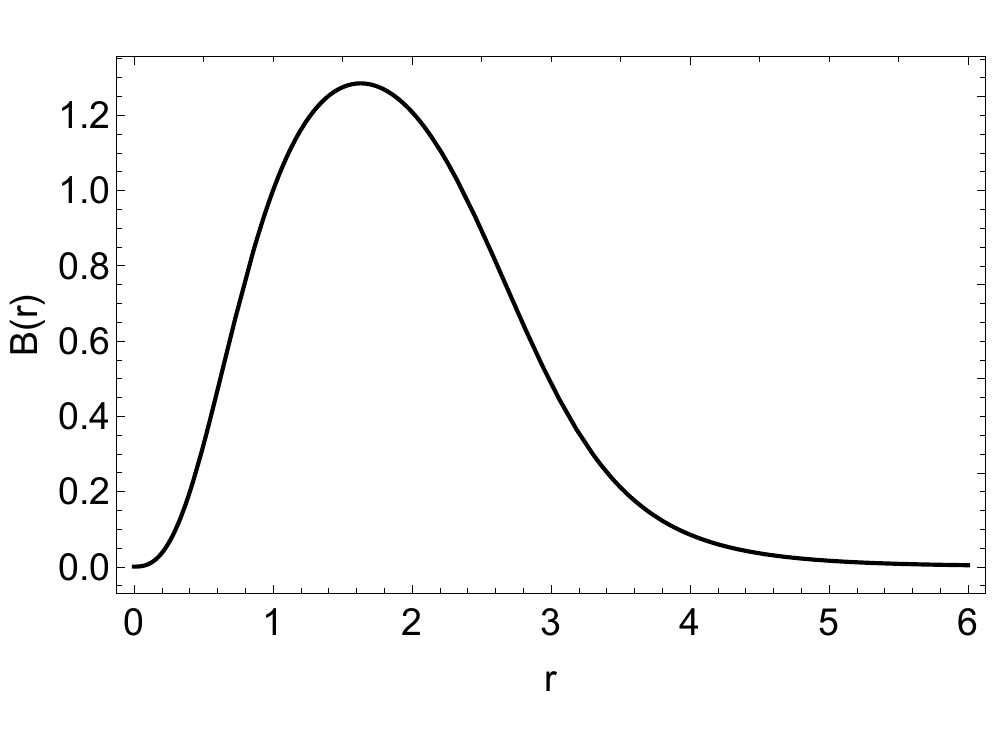}}
    \subfigure[Planar magnetic configuration.]{\includegraphics[width=6.5cm,height=6cm]{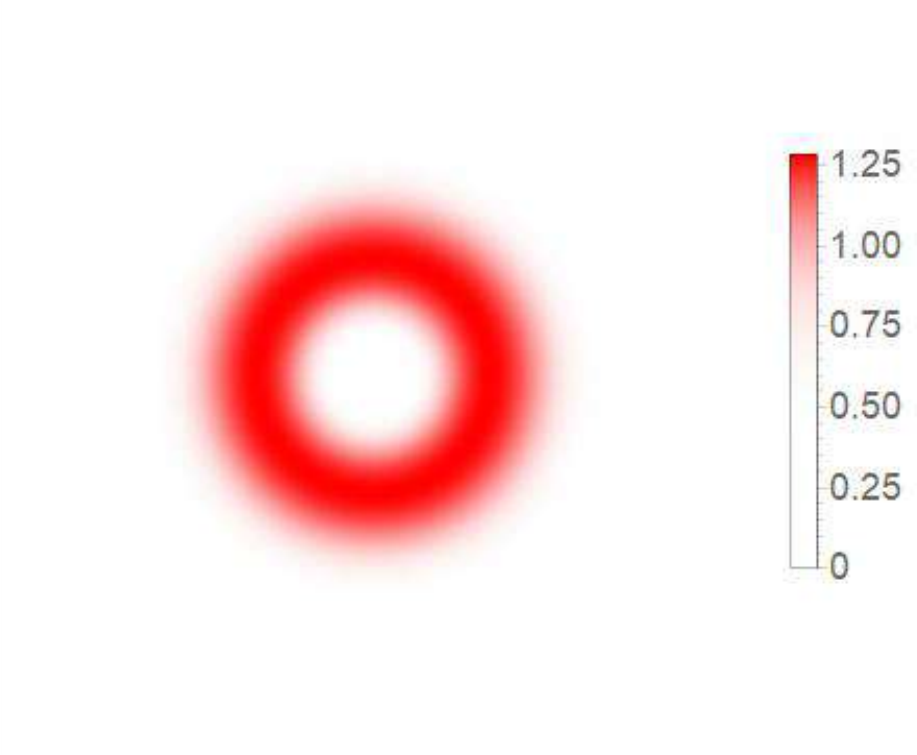}}
    \caption{Magnetic field $B(r)$ vs. $r$ with $g=0.025$, $e=1$ and unit winding number.}
    \label{Fig2}
\end{figure}

We obtained the numerical solutions for the field variables $f(r)$ and $a(r)$ in the $\CP^1$-Maxwell model. Let us proceed to investigating the numerical solutions for the magnetic field. For this purpose, one recalls that the magnetic field is $B = F_{12} = -\frac{a'(r)}{e r}$. Accordingly, by employing the numerical solutions displayed in Fig. \ref{Fig1}, the corresponding numerical solutions for the magnetic field are obtained and presented in Fig. \ref{Fig2}.

Figures \ref{Fig2}(a) and \ref{Fig2}(b) display the radial profile of the magnetic field associated with the BPS solutions in the $\CP^1$-Maxwell model, providing a direct physical characterization of the magnetic flux distribution of these lump-like configurations. One observes that the magnetic field is strictly localized, reaching its maximum value in the vicinity of the core and decaying rapidly as the radial distance increases, thereby confirming the confined nature of the magnetic flux.
This behavior follows directly from the second BPS equation, which establishes a rigid functional relation between the magnetic field and the modulus of the $\CP^1$ field. Particularly, since $f(r)\to 0$ both in the limits $r\to 0$ and $r\to \infty$, the magnetic field also vanishes in these limits. Consequently, one finds a magnetic flux concentrated at an intermediate region where $f(r)$ attains its maximum value, resulting in a lump with a ring-like profile and reduced intensity at the center. Figure \ref{Fig2}(b), which emphasizes the planar behavior, further highlights the nonsingular nature of the solution, showing that $B(r)$ remains finite throughout the entire radial domain and exhibits no divergences at the core.

\begin{figure}[!ht]
    \centering
    \subfigure[]{\includegraphics[width=7cm,height=6cm]{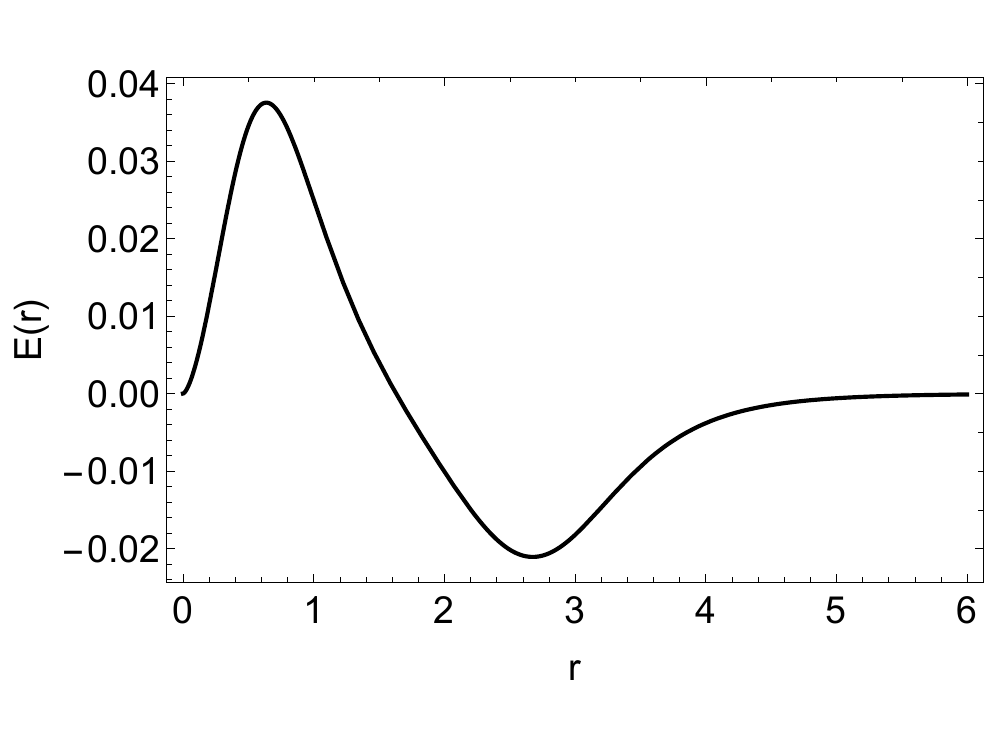}}
    \subfigure[]{\includegraphics[width=7cm,height=6cm]{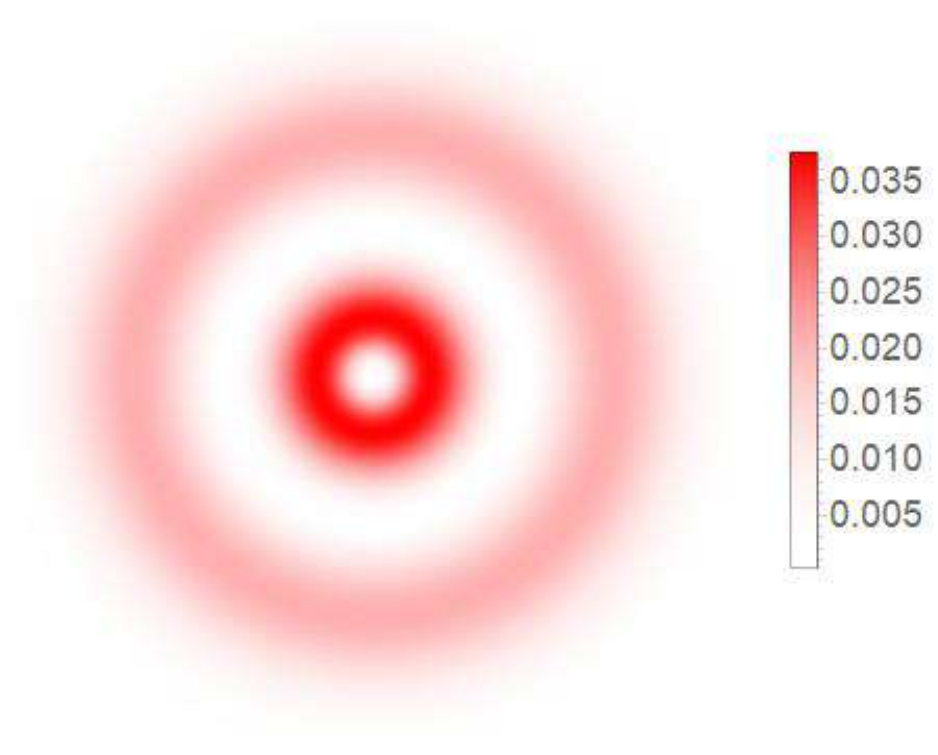}}
    \caption{(a) Electric field profile $\mathrm{E}(r)$ vs. $r$, with $e=1$ and $g=0.025$. (b) Planar distribution of the electric field modulus for $e=1$ and $g=0.025$.}
    \label{Figx}
\end{figure}

Meanwhile, taking into account that $\mathrm{E}(r)=-A'_0(r)$, we analyze the behavior of the electric field. In this case, the electric field profile is displayed in Figures \ref{Figx}(a) and \ref{Figx}(b).
Figures \ref{Figx}(a) and \ref{Figx}(b) show that the induced electric sector is nontrivial, regular, and spatially localized. The radial profile of the electric field exhibits a positive maximum near the inner region of the solution, changes sign within an intermediate range, and asymptotically decays to zero, indicating a radial redistribution of the electric field without singularities at the core or any relevant long-range contribution. In turn, the planar distribution of $\vert\mathrm{E}\vert$ confirms the circular symmetry of the configuration and shows that the electric intensity remains confined around the solitonic structure, with modulation in concentric rings and rapid spatial attenuation, which is consistent with a magnetized and electrically polarized BPS solution.

\begin{figure}[!ht]
    \centering
    \includegraphics[width=7cm,height=6cm]{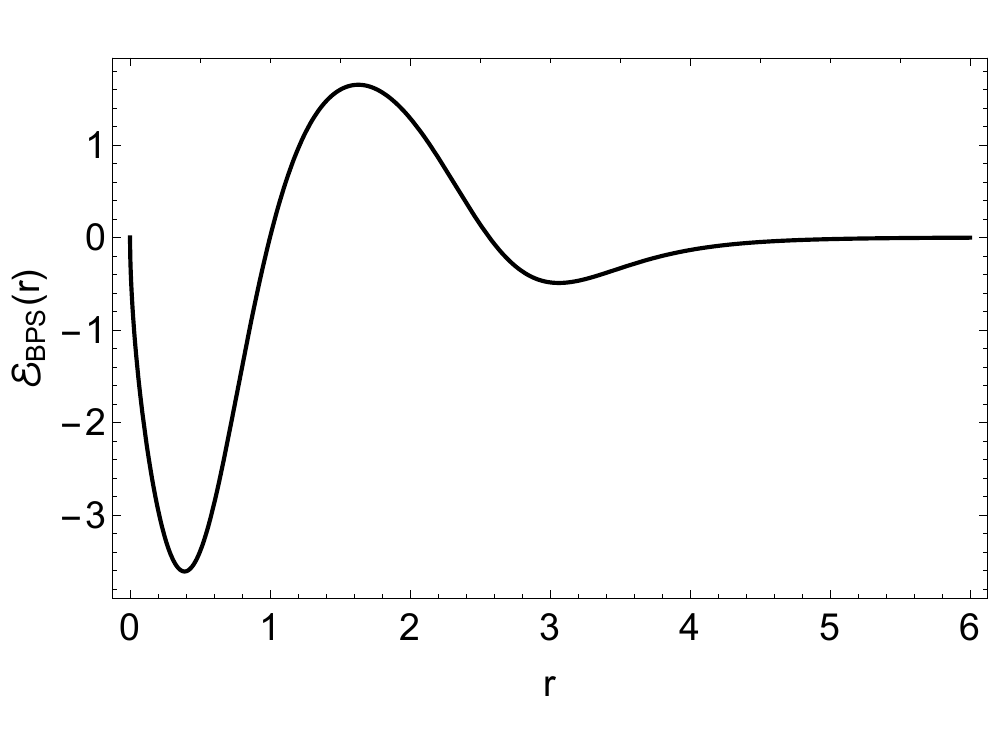}
    \caption{BPS energy density $ \mathcal{E}_{\mathrm{BPS}}(r)$ vs. $r$ with $e=1$ and $g=0.025$.}
    \label{Fig3}
\end{figure}

To conclude, let us consider the BPS energy density from Eq. \eqref{Eq19}, together with the solutions of the field variables shown in Fig. \ref{Fig1}. Taking into account the numerical approach, one obtains the BPS energy density; see Fig. \ref{Fig3}.

Figure \ref{Fig3} display the BPS energy density $\mathcal{E}_{\mathrm{BPS}}(r)$ in terms of the radial coordinate, providing a complementary and energetically more refined characterization of the internal structure of the lump. The profiles confirm that the energy density is finite, positive, and strongly localized, with a maximum located approximately in the same region where the magnetic field reaches its highest intensity, reflecting the interdependence between the scalar and gauge sectors in the self-dual regime. Near the origin, the energy density is null. Meanwhile, in the asymptotic limit, it decays rapidly, ensuring the convergence of the total energy. Furthermore, one notes the absence of singularities and the smooth radial decay, indicating that saturation of the BPS bound implies that the total energy depends exclusively on the boundary values of the fields, rather than on their local fluctuations.

\section{Summary and Conclusion}

In this work, we investigated self-dual radially symmetric configurations in the generalized $\CP^1$ model coupled to a Maxwell--Chern--Simons gauge field through nonminimal interactions. Starting from the nonlinear $O(3)$ sigma model, we established its classical mapping into the $\CP^1$ formulation, emphasizing the emergence of a local $U(1)$ gauge symmetry and the central role played by the Fubini--Study geometry of the target space.

The generalized gauge sector, characterized by the simultaneous presence of the Chern--Simons term and the Pauli-like nonminimal coupling, modifies the effective gauge connection and changes the structure of the static regime. In particular, Gauss's law shows that the electric sector cannot, in general, be consistently suppressed. As a consequence, the self-dual configurations supported by the model are not merely magnetized, but also electrically polarized, with electric and magnetic fields intrinsically coupled already at the level of the field equations.

By adopting a radially symmetric ansatz, we showed that the topological boundary conditions lead to a quantized magnetic flux determined exclusively by the asymptotic behavior of the gauge field. This quantization remains valid even in the presence of both the CS contribution and the nonminimal interaction, showing that the topological content of the solutions is robust under the generalization of the gauge sector.

The implementation of the BPS approach allowed us to identify the self-interaction potential required for the existence of self-dual configurations. In this framework, the energy can be reorganized into a sum of positive-definite terms plus a boundary contribution, yielding a consistent set of first-order BPS equations. The resulting self-dual structure makes explicit the interplay among the scalar profile, the magnetic field, and the induced electric sector, while ensuring the classical stability of the solutions.

The asymptotic analysis revealed that finite-energy solutions necessarily require the $\CP^1$ scalar field to vanish at spatial infinity. Therefore, the admissible self-dual configurations are lump-like rather than Nielsen--Olesen-like vortices, since they do not interpolate between distinct vacua. 

The numerical solutions corroborate the analytical results. The scalar and gauge profiles are regular over the entire radial domain and satisfy the required boundary conditions. The magnetic field is spatially localized and concentrated in an intermediate radial region, whereas the electric field is nontrivial, regular, and also localized, exhibiting a sign change before vanishing asymptotically. Altogether, these results characterize the solutions as magnetized and electrically polarized BPS lumps with no singular behavior either at the core or at infinity.

In summary, the generalized $\CP^1$--Maxwell--CS model with nonminimal coupling supports regular self-dual solitonic configurations whose magnetic flux is quantized, whose electric sector is dynamically induced, and whose existence is rigidly controlled by the Fubini--Study geometry of the target space. These results extend the standard $\CP^1$--Maxwell scenario and show that the combined effects of the Chern--Simons term and the anomalous magnetic moment interaction provide a natural mechanism for the emergence of electrically polarized BPS lumps.

\section*{ACKNOWLEDGMENT}

F. C. E. Lima and I. B. Cunha would like to express their sincere gratitude to the Conselho Nacional de Desenvolvimento Científico e Tecnológico (CNPq) and Fundação de Amparo \`{a} Pesquisa do Estado de S\~{a}o Paulo (FAPESP) for their valuable support. F. C. E. Lima is supported, respectively, for grants No. 2025/05176-7 (FAPESP) and 171048/2023-7 (CNPq). A. V. has been funded by FONDECYT Iniciaci\'on No. 11261883.

\section*{DATA AVAILABILITY}

No data was used for the research described in this article.

\bibliographystyle{apsrev4-2}
\bibliography{refs}

\end{document}